# Further exploration of the machine-learning-based nuclear mass table*

LIU Yaqi[1,2,3], LI Zhilong[2,4], WANG Yongjia[2,*], LI Qingfeng[2,*], MA Chunwang[1,3]

1. College of Physics, Henan Normal University, Xinxiang 453007, China
2. School of Science, Huzhou University, Huzhou 313000, China
3. Institute of Nuclear Science and Technology, Henan Academy of Sciences, Zhengzhou 450046, China
4. China Institute of Atomic Energy, Beijing 102413, China

**Abstract**

The mass of the atomic nucleus, as one of the fundamental physical quantities of the atomic nucleus, plays an important role in understanding and researching the structure of the atomic nucleus and nuclear reactions, and the basic interactions between nucleons. However, accurately predicting the mass of nuclei far from the β stability line remains a huge challenge. Based on the machine-learning-refined mass model, we investigate the newly measured atomic nucleus masses since 2022, along with the residual proton-neutron interaction ($\delta V_{\mathrm{pn}}$) and the $\alpha$-decay energy of heavy nucleus. It is found that: 1) For the 23 newly measured atomic nuclei, the root mean square deviations obtained by the machine-learning-refined mass models are between 0.51 and 0.58 MeV, which are significantly lower than 3.275, 1.058, 0.752, and 0.785 MeV given by the liquid droplet model (LDM), Weizsäcker-Skyrme-4 (WS4), finite-range droplet model (FRDM), and Duflo-Zucker (DZ), respectively. 2) The $\delta V_{\mathrm{pn}}$ of the atomic nucleus with $N = Z$ obtained from machine-learning-refined mass models is consistent with the latest experimental data. 3) The root mean square deviations of the $\alpha$-decay energy of heavy nuclei obtained from the machine-learning-refined mass models have also been significantly reduced. Furthermore, by employing the Bayesian model average approach to combine the results from different machine-learning-refined mass models, we obtain more accurate predictions. These findings demonstrate that such models have good extrapolation capabilities and provide useful insight for further research. The datasets presented in this paper are openly available at https://doi.org/10.57760/sciencedb.j00213.00246.

* The paper is an English translated version of the original Chinese paper published in Acta Physica Sinica. Please cite the paper as: **LIU Yaqi, LI Zhilong, WANG Yongjia, LI Qingfeng, MA Chunwang, Further exploration of the machine-learning-based nuclear mass table.** *Acta Phys. Sin.*, 2026, 75(2): 020107. DOI: 10.7498/aps.75.20251526

# 1 Introduction

Machine learning, as an efficient technique for data processing and analysis, is transforming human production and lifestyles. It has also influenced many scientific research fields, including physics, establishing data-driven scientific research as the fourth paradigm of science following experimental, theoretical, and computational approaches. Leveraging its capacity to process and analyze large, complex, high-dimensional, and deeply correlated datasets, the integration of machine learning with nuclear physics and nuclear data has yielded valuable research outcomes[1-9]. For instance, in nuclear structure research, machine learning methods are widely applied to study nuclear masses[10-16], neutron skin thickness[17,18], charge radii[19-23], and nuclear energy levels[24-26]. In the fields of nuclear decay and reactions, these methods are extensively used to investigate nuclear half-lives[27-31], nuclear matter phase transitions[32-34], heavy-ion fusion cross-sections[35,36], fission yields and radioactive nuclide migration[37-39], projectile fragmentation cross-sections[40], and spallation cross-sections[41].

Mass is the most fundamental property of atomic nuclei. It holds significant importance in studying nuclear structure and reactions, nucleon-nucleon interactions, nuclear decay and fission processes, and the synthesis and detection of superheavy nuclei. Since atomic nuclei are complex quantum many-body systems composed of numerous strongly interacting nucleons with significant correlation effects, accurately predicting nuclear masses, especially for nuclei far from the $\beta$-stability line, remains a major challenge. Nuclear mass models are generally categorized into two types: global and local. Local mass formulas consider specific relationships between the masses of adjacent nuclides. They cancel out nucleon interactions through concise algebraic forms, thereby enabling mass predictions for certain nuclei[42]. Global mass models can be further subdivided into macroscopic, macroscopic-microscopic, and microscopic models. Macroscopic-microscopic models typically evolve from the famous Bethe-Weizsäcker (BW) nuclear mass formula by incorporating various microscopic corrections. Examples include the Finite Range Droplet Model (FRDM)[43], the Duflo-Zuker (DZ) model[44], and the Weizsäcker-Skyrme (WS) model[45]. Common microscopic models include the Relativistic Mean-Field (RMF) model based on density functional theory[46,47], the Skyrme-Hartree-Fock-Bogoliubov model[48], the Gogny-Hartree-Fock-Bogoliubov model[49], and the Deformed Relativistic Hartree-Bogoliubov theory in Continuum (DRHBc)[50]. Microscopic calculations based on first principles can currently only reliably explore nuclei with small mass numbers. Other calculations based on various effective theories often depend on the selection of model parameters. Therefore, it is essential to develop high-precision methods for calculating nuclear masses.

The masses of nearly 3,000 nuclides have been measured. These data provide a foundation for constructing machine learning-based nuclear mass tables[51-54]. Reference [55] utilized 2,408 nuclear masses from $^{16}$O to $^{270}$Ds taken from AME2016[56] as the training set (80%) and test set (20%). The study considered ten features, including proton number, neutron number, and distance to the nearest magic number. It employed the Light Gradient Boosting Machine (LightGBM) to optimize the mass table for nuclei with proton and neutron numbers greater than 8. The results indicated that the root mean square deviations (RMSD) between the predicted and experimental values for the Liquid-Drop Model (LDM), Duflo-Zuker (DZ) model, Weizsäcker-Skyrme-4 (WS4) model, and Finite Range Droplet Model (FRDM) on the test set were 0.234, 0.213, 0.170, and 0.222 MeV, respectively. Compared to the RMSD values of the uncorrected mass models, these represented reductions of approximately 90%, 65%, 40%, and 60%, respectively[55].

Between 2022 and 2025, the measurement of more than 20 new nuclear masses further validates the reliability of machine learning-optimized nuclear mass tables. Superheavy nuclei are a current hotspot in nuclear physics research, and $\alpha$ decay is a crucial method for identifying them. In recent years, the accumulation of experimental data on $\alpha$ decay energies ($Q_\alpha$) for heavy and superheavy nuclei has provided new means to systematically evaluate the reliability of various nuclear mass tables. This is achieved through machine learning methods, such as Radial Basis Function (RBF)[30] and Extreme Gradient Boosting (XGBoost)[57], to predict $Q_\alpha$ values. Furthermore, the residual proton-neutron interaction ($\delta V_{pn}$) describes the relationship between the binding energy of a nucleus and its neighbors[58]. This interaction is inseparable from nuclear mass and serves as an important tool for verifying machine learning-refined nuclear mass tables. To verify the reliability and extrapolation feasibility of machine learning methods, this study investigates the performance of machine learning algorithms in predicting the masses of recently measured nuclei, residual proton-neutron interactions, and α decay energies.

# 2 Methods

## 2.1 Machine Learning Algorithms

LightGBM (Light Gradient Boosting Machine) is a decision tree algorithm based on the histogram concept. It introduces several improvements and optimizations to the traditional Gradient Boosting Decision Tree (GBDT) algorithm. Unlike GBDT, which uses an inefficient level-wise tree growth strategy, LightGBM employs a leaf-wise growth algorithm with depth constraints. For the same number of splits, leaf-wise growth reduces error and achieves higher precision. LightGBM uses a histogram-based decision tree algorithm that discretizes continuous feature values into fixed bins. This approach not only improves computational efficiency but also reduces memory usage. To prevent overfitting, LightGBM provides L1 and L2 regularization, allowing users to balance model complexity and fitting performance by adjusting regularization terms. Additionally, LightGBM supports multi-threaded parallel computing and distributed training. Its efficient feature splitting selects optimal split points rather than exhaustively searching all possible splits for all features, making it particularly suitable for large-scale datasets. These advantages enable LightGBM to achieve higher precision, faster computation speeds, and the capacity to handle large datasets. During the "gradient boosting" process, the target predicted in each iteration is the residual between the previous iteration's result and the final target. This algorithm is applicable to various feature types and offers advantages such as fast execution and strong interpretability.

We use LightGBM to train on the residual $\delta(Z,A)$ between the experimental binding energy and the binding energy given by theoretical models, thereby predicting binding energies. The residual is defined as $\delta(Z,A)=B_{\mathrm{th}}(Z,A)-B_{\mathrm{exp}}(Z,A)$, where $B_{\mathrm{th}}$ represents the theoretical model value and $B_{\mathrm{exp}}$ represents the experimental value. The four models used are LDM, DZ, WS4, and FRDM, respectively. The binding energy of unknown nuclei predicted by machine learning is obtained via $B_{\mathrm{LightGBM}}(Z,A)=B_{\mathrm{th}}(Z,A)-\delta(Z,A)$. Furthermore, this study employs Bayesian Model Averaging (BMA) to weight the four machine learning-optimized mass models and provide BMA predictions. BMA is a technique that combines multiple models for prediction through Bayesian inference. It performs a weighted average of all candidate models to obtain the final prediction, thereby reducing errors potentially introduced by individual models. Specifically, it calculates the posterior probability of each model using Bayes' theorem to determine weights. The predictions of these models are then weighted and averaged according to these weights to yield a more accurate prediction. This algorithm is suitable for multi-model ensemble learning and regression analysis, offering advantages such as reduced overfitting, improved stability, and strong adaptability. For a detailed introduction to BMA, see references [59,60].

### 2.2 Data and Input Features

This study uses the binding energies of 2,408 nuclei between $^{16}$O and $^{270}$Ds from AME2016 as the training set. The test dataset consists of 66 new binding energies added in AME2020. The $\alpha$ decay energies are sourced from references [11,61]. The ten input features selected are: mass number $A$, proton number $Z$, neutron number $N$, neutron-to-proton ratio $N/Z$, theoretical value from LDM $B_{\mathrm{LDM}}$, pairing effect dependency $N_{\mathrm{pair}}$=0,1,2 (odd-odd, odd-even, even-even nuclei), last proton shell $Z_m=1,2,\cdots$ ($8\leqslant Z<20, 20\leqslant Z<28,\cdots$), last neutron shell $N_m=1,2,\cdots$ ($8\leqslant N<20, 20\leqslant N<28,\cdots$), distance between proton number and the nearest magic number $|Z-m|$ ($m\in\{8,20,28,50,82,126\}$), and distance between neutron number and the nearest magic number $|N-m|$ ($\mathrm{m}\in\{8,20,28,50,82,126\}$). To ensure the stability and validity of model evaluation, key hyperparameters were systematically configured, as detailed in Table 1. Other hyperparameters in the model have a minor impact on performance and thus retain default settings. We randomly selected 80% of the nuclear data from AME2016 as the training set. After optimizing the four mass models, we provided masses for unknown nuclides; relevant data are available in the supplementary materials of reference [55]. BMA calculates the posterior probability of each model using Bayes' theorem and combines the predictions of these models through a weighted average to obtain a more accurate prediction. Reference [62] used BMA to weight the Finite Range Droplet Model 2012 (FRDM12) and Hartree-Fock-Bogoliubov-31 (HFB-31) mass models, resulting in a mass table with higher precision.

**Table 1** LightGBM hyperparameter configuration.

| Parameter | Description | Value |
|---|---|---|
| num_leaves | Maximum number of leaves per tree | 10 |
| min_data_in_leaf | Minimum number of data points in each leaf | 20 |
| max_depth | Maximum depth of each tree (-1 indicates no limit) | -1 |
| learning_rate | Learning rate | 0.1 |
| lambda_l1 | L1 regularization term | 0.1 |
| num_boost_round | Number of boosting iterations (number of trees) | 20000 |
| verbose_eval | Logging frequency during training | 50 |
| early_stopping_rounds | Rounds with no improvement to stop early | 100 |

This study employs the root mean square deviation to evaluate the agreement between the mass table and experimental data. The calculation method is given by the following equation:

$$\sqrt{\sigma^2} = \sqrt{\frac{1}{\tilde{n}}\sum_{p=1}^{\tilde{n}} (Q^{\mathrm{Expt.p}} - Q^{\mathrm{Theo.p}})^2}, \quad (1)$$

where $Q^{\mathrm{Expt.p}}$ denotes the experimental value and $Q^{\mathrm{Theo.p}}$ denotes the theoretical value.

# 3 Results and Discussion

## 3.1 Predictive Capability for Newly Measured Nuclear Masses

Since 2022, experimental measurements have determined the masses of 23 new nuclei. Figure 1 displays the deviations between the binding energies calculated using the machine learning-optimized mass tables and the experimental values. Figure 1 indicates that the machine learning corrections improved the predictions for these 23 nuclei to varying degrees across different models. For instance, in the Liquid Drop Model (LDM), $^{95}$Ag and $^{152}$La showed significant improvements. Similarly, improvements were observed for $^{155}$Ce and $^{154}$Pr in the DZ model, $^{31}$Ar and $^{152}$Ce in the WS4 model, and $^{37}$Sc and $^{30}$F in the FRDM. Compared to the pre-optimization results, the nuclear binding energies after optimization align much more closely with experimental values, specifically for $^{95}$Ag, $^{152}$La and $^{37}$Sc *wi*th the deviations from experimental values decreased by 97.5%, 98.4%, and 99.0%, respectively. The nucleus $^{95}$Ag lies in the vicinity of the double magic numbers $N = Z = 50$. Nuclides around this region are close to the proton drip line and exhibit characteristics of N≈Z. Their energy level structures are highly sensitive to the subtle balance of nuclear forces, making them more prone to displaying pure quantum shell structure effects. Investigating nuclei near the $Z = N = 50$ double magic region helps verify the robustness of shell closure under conditions of extreme isospin asymmetry. It also aids in understanding the resulting binding energy enhancement effects and deepens the comprehension of the origins of nuclear shell structures. Crucially, data for such nuclei are vital for validating and developing the "shell effect" in nuclear models. This provides key theoretical and experimental foundations for further constructing more precise nuclear mass models[63]. Taking the liquid drop model as an example, the traditional LDM includes only volume, surface, Coulomb, and pairing energy terms, excluding shell effects. In this machine learning study, we incorporated physical quantities related to shell effects, such as the difference between the proton number and the nearest magic number, denoted as $v_p$, and the difference between the neutron number and the nearest magic number, denoted as $v_n$. These additions significantly improved the prediction of nuclear binding energies. Furthermore, the RMSD for the LDM, DZ, WS4, and FRDM mass tables optimized using LightGBM are 0.570, 0.558, 0.581, and 0.512 MeV, respectively. These values are lower than the pre-optimization RMSD of 3.275, 1.058, 0.752, and 0.785 MeV. This demonstrates that the machine learning-optimized mass tables maintain high reliability when predicting masses of nuclides in unknown nuclear regions. Additionally, these RMSD values are higher than the results reported in our previous study [55]. This discrepancy arises because the nuclei investigated in this work are further from the training set than those in the prior study. To further evaluate performance, this study employed the Bayesian Model Averaging (BMA) method to integrate the four machine learning-optimized models, reducing the RMSD to 0.468 MeV. This value outperforms results from existing methods such as Ensemble Learning and Model Averaging (ELMA)[64], Sequential Least Squares Programming (SLSQP)[65], and Bayesian Neural Networks (BNN)[52]. These findings verify the strong generalization capability of the machine learning algorithm proposed in this paper. Relevant data are presented in Table 2.

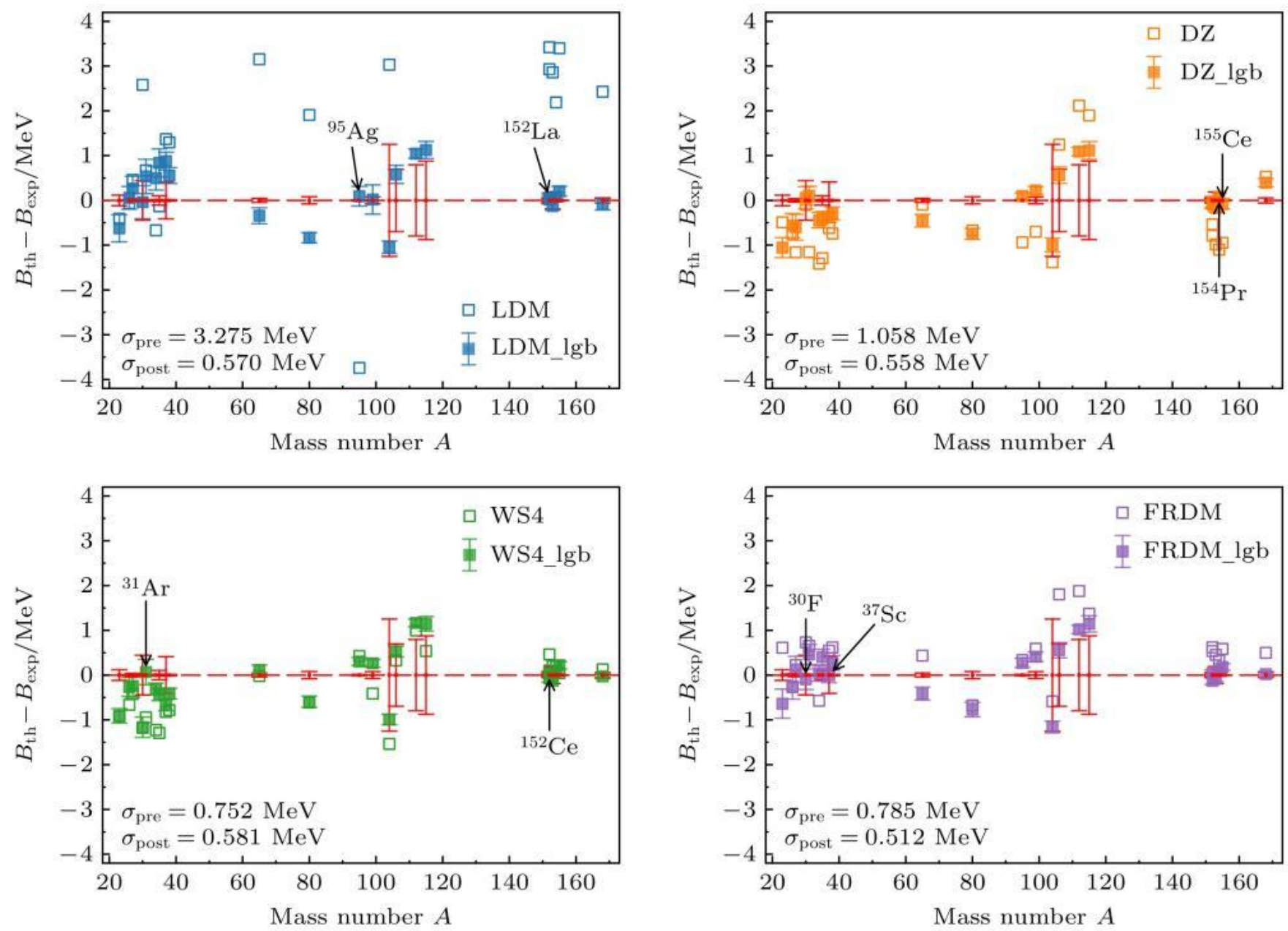

**Fig. 1 Difference between the theoretical and the experimental binding energies (red horizontal line) obtained using the LDM, DZ, WS4, and FRDM (open squares) and LightGBM-refined mass models (solid squares). The results show the differences between the theoretical values and the experimental values of the binding energies of 23 newly measured nuclides. $\sigma_{pre}$ and $\sigma_{post}$ denote the RMSD of the original and LightGBM-refined mass models on the newly measured nuclei, respectively. The error of the predictions obtained using the LightGBM-refined mass model is the standard deviation of the predicted binding energy. It is obtained by running LightGBM 500 times with randomly splitting**

**AME2016 data into training and test sets with a ratio of 4:1.**

**Table 2** A comparison is conducted of the predicted nuclear binding energies and RMSD on the test set. The results from three established approaches—ELMA, BNN and SLSQP — are systematically compared with those obtained using BMA, thus verifying the performance differences among the various methods (Unit: MeV)

| Z | A | EXP. | BMA (this work) | ELMA[64] | BNN[52] | SLSQP[65] | Reference |
|---|---|---|---|---|---|---|---|
| 9 | 30 | 186.399 | 186.361 | - | 186.672 | 185.995 | [66] |
| 14 | 23 | 151.148 | 150.381 | 151.085 | 150.302 | 150.494 | [67] |
| 15 | 26 | 187.118 | 187.043 | - | 187.196 | 184.879 | [67] |
| 16 | 27 | 187.987 | 187.987 | 188.253 | 187.902 | 187.165 | [67] |
| 18 | 31 | 224.835 | 224.977 | 224.973 | 225.230 | 224.764 | [67] |
| 19 | 34 | 261.043 | 261.043 | - | 260.911 | 260.219 | [68] |
| 20 | 35 | 262.068 | 262.068 | 262.098 | 262.223 | 263.655 | [69] |
| 21 | 37 | 278.705 | 278.694 | 278.389 | 279.114 | 280.240 | [68] |
| 21 | 38 | 294.932 | 294.932 | - | 294.832 | 296.323 | [68] |
| 24 | 65 | 534.055 | 533.934 | 533.901 | 534.267 | 534.566 | [70] |
| 39 | 104 | 864.099 | 863.056 | 862.352 | 862.650 | 863.113 | [71] |
| 40 | 80 | 669.528 | 668.803 | 669.026 | 668.824 | 668.710 | [72] |
| 40 | 106 | 882.356 | 882.912 | 882.482 | 883.037 | 883.728 | [71] |
| 42 | 112 | 927.662 | 928.739 | 928.188 | 928.767 | 929.627 | [71] |
| 43 | 115 | 949.638 | 950.770 | 950.182 | 950.490 | 951.570 | [71] |
| 47 | 95 | 789.734 | 789.878 | 790.124 | 789.819 | 786.780 | [63] |
| 49 | 99 | 822.140 | 822.208 | 822.779 | 821.989 | 816.991 | [73] |
| 57 | 152 | 1231.413 | 1231.383 | 1231.143 | 1231.716 | 1231.526 | [74] |
| 58 | 152 | 1240.329 | 1240.347 | 1240.087 | 1240.405 | 1240.048 | [75] |
| 58 | 153 | 1244.347 | 1244.235 | 1244.060 | 1244.471 | 1243.896 | [75] |
| 58 | 155 | 1253.265 | 1253.338 | 1252.975 | 1253.511 | 1252.757 | [74] |
| 59 | 154 | 1254.807 | 1254.896 | 1254.582 | 1254.884 | 1253.958 | [75] |
| 64 | 168 | 1354.051 | 1354.051 | 1354.027 | 1354.026 | 1353.073 | [74] |
| RMSD | | | 0.468 | 0.514 | 0.529 | 1.650 | |

## 3.2 Residual Proton-Neutron Interaction $\delta V_{pn}$

The residual proton-neutron interaction plays a crucial role in atomic nuclei. It significantly influences nuclear binding energy, level structure, nuclear shape, and nuclear reaction processes. Specifically, this interaction determines nuclear stability, the distribution of excited and ground states, as well as cross-sections and product distributions in nuclear reactions. Investigating $\delta V_{pn}$ holds significant scientific value for improving nuclear structure models, such as the nuclear shell model and collective model, and for understanding nucleosynthesis processes in stars[76]. The residual proton-neutron interaction is defined as the double difference of binding energies among four adjacent nuclei, which can be expressed as[58]

$$\delta V_{pn}^{ee}(Z,N)=\frac{1}{4}[B(Z,N)-B(Z,N-2) \\ -B(Z-2,N)+B(Z-2,N-2)]. \quad (2)$$

The above equation requires that the number of protons and neutrons in the nucleus be equal and both even ($\delta V_{pn}^{ee}$). For odd-odd nuclei, it requires that the number of protons and neutrons be equal and both odd ($\delta V_{pn}^{oo}$), which is defined as

$$\delta V_{pn}^{oo}(Z,N)=[B(Z,N)-B(Z,N-1) \\ -B(Z-1,N)+B(Z-1,N-1)]. \quad (3)$$

Recently, the Heavy Ion Research Facility in Lanzhou employed isotope mass spectrometry to measure the masses of nuclei such as $^{62}$Ge , $^{64}$As, and $^{66}$Se. These new measurements allow for the calculation of $\delta V_{pn}$ for nuclei with $Z$=$N$ . For even-even nuclei (odd-odd nuclei) with $Z \geqslant 28$, the magnitude decreases (increases) as the mass number $A$ increases[58]. Figure 2 presents these results. The $\delta V_{pn}$ values for odd-odd nuclei predicted by the four mass models are significantly lower than the experimental values. In contrast, the results for even-even nuclei align closely with experimental data. For the anomalous bifurcation of $\delta V_{pn}$ observed in Z=N nuclei, previous studies have suggested that three-body forces play an important role in $\delta V_{pn}$[58] . Reference [77] successfully reproduced this anomalous behavior using a relativistic many-body theory. This method uniformly considers neutron-neutron, proton-proton, and neutron-proton pairing correlations. It strictly handles blocking effects and maintains particle number conservation. The study attributes this phenomenon to enhanced proton-neutron pairing correlations in odd-odd nuclei. Figure 2 shows that results from mass models optimized via machine learning align more closely with experimental values. Notably, the binding energies for $^{70}$Br and $^{65}$As in the training set are 589.03752 MeV and 545.755 MeV, respectively. Using these values, the calculated $\delta V_{pn}$ for $^{70}$Br and $^{66}$As are 2.923 MeV and 2.925 MeV. However, the $\delta V_{pn}$ values for $^{70}$Br and $^{66}$As in the figure are derived from binding energies (589.630 MeV and 546.397 MeV) reported in Reference [58]. These yield $\delta V_{pn}$ values of 3.515 MeV and 3.066 MeV. Consequently, the $\delta V_{pn}$ values for these two nuclei from the machine-learning-optimized mass table are lower than the experimental values. Calculating $\delta V_{pn}$ for $^{64}$Ge , $^{68}$Se , $^{72}$Kr , and $^{74}$Rb requires the masses of $^{62}$Ge , $^{66}$Se , $^{70}$Kr *and* $^{73}$Rb. These nuclear data were not included in the training set. This demonstrates the strong optimization capability of machine learning methods for nuclear mass models. Figure 3 displays the results obtained using the Bayesian Model Averaging (BMA) method. The BMA results align more closely with experimental values. For mass numbers greater than 76, no experimental $\delta V_{pn}$ data are currently available. Results calculated from evaluation values in AME2020 indicate a peak at $^{82}$Nb . The machine-learning-optimized LDM, WS4, and DZ models also exhibit distinct peaks at this point. The machine-learning-optimized FRDM model and the BMA results show weaker peaks. Current experimental data are insufficient to constrain model parameters in these regions. Future measurements of these nuclei will provide further opportunities to validate nuclear mass tables.

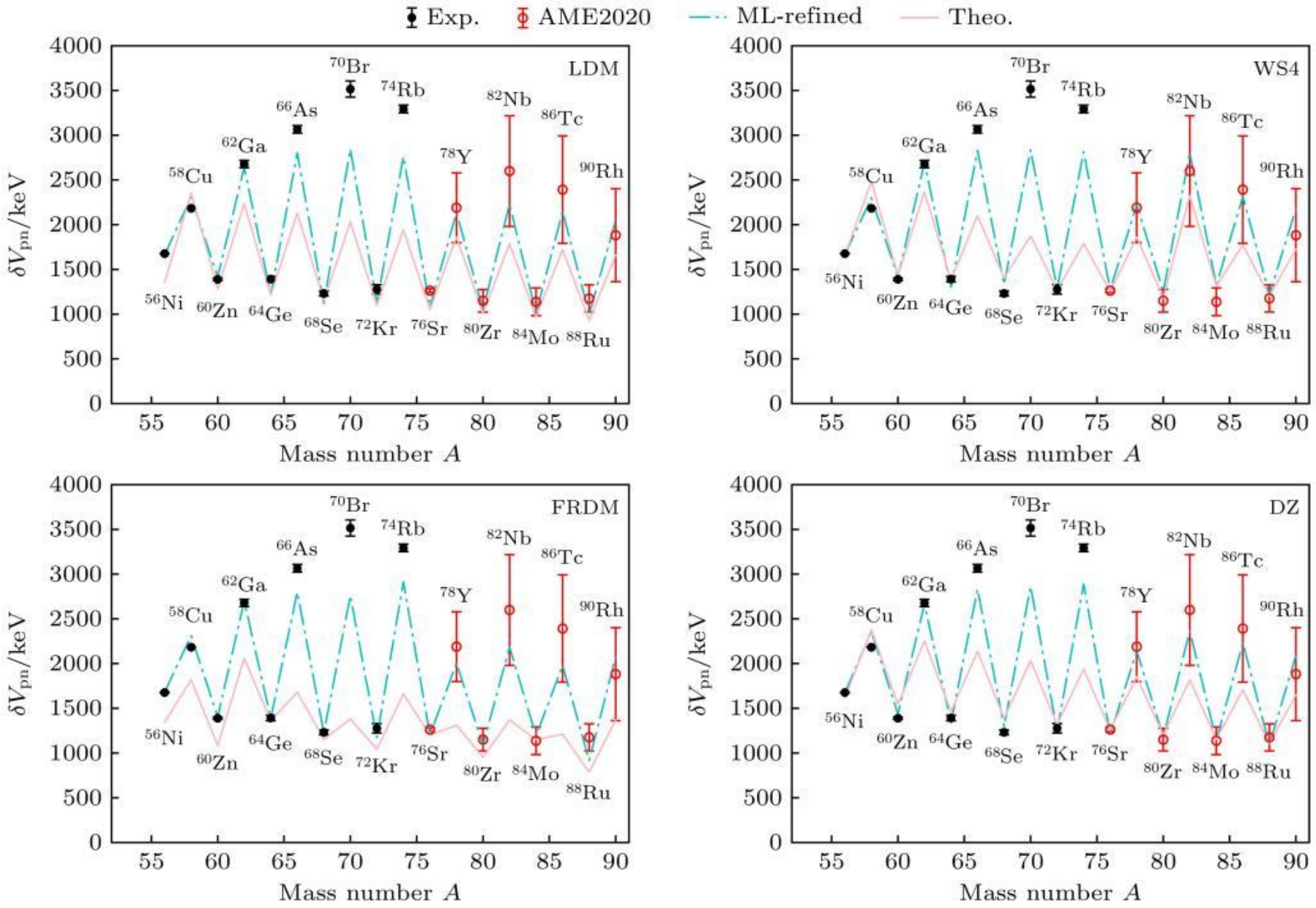

**Fig. 2 Experimental $\delta V_{pn}$ for $Z = N$ nuclei beyond $A = 56$ and comparison to different mass model predictions. Solid black dots represent experimental value, the data taken from literature [51,58]. The red hollow circles represent the results obtained from the AME2020 evaluation data. The solid lines represent the results given by the four mass models, while the dashed lines indicate the results provided by the four models after machine learning optimization.**

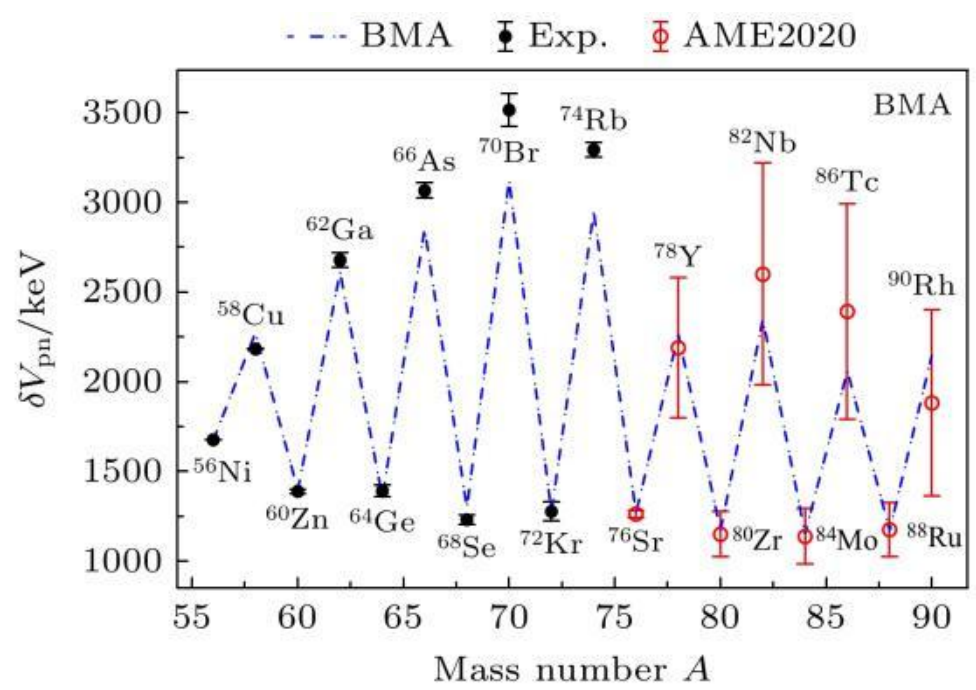

**Fig. 3 The $\delta V_{pn}$ of nuclei with $A \geqslant 56$ and $Z=N$ is compared with the results of BMA. The black solid dots represent the experimental values, the red hollow circles represent the results obtained from the AME2020 evaluation data, and the blue dashed lines represent the prediction through BMA. The experimental data used in this figure is consistent with that in Fig. 2.**

### 3.3 Alpha Decay Energy of Atomic Nuclei

Accurately predicting the $\alpha$ decay energy of superheavy nuclei is crucial for their laboratory synthesis and detection. The $\alpha$ decay energies of nuclei can be derived from nuclear binding energies. *This study calculates the* α decay energies for elements with atomic numbers 89-120 using an optimized mass table. The relevant data are provided in the supplementary spreadsheet[78]. Table 3 lists the RMSD between the α decay energies derived from various mass models and the experimental values. The results indicate that the RMSD values from the machine-learning-optimized mass tables are significantly lower than those from the pre-optimization models. Specifically, for nuclides with $89 \leqslant Z \leqslant 102$, the RMSD of the α decay energy predicted by the optimized mass tables can fall below 100 keV. This precision holds substantial value for experimental studies of nuclides in this region. For nuclides in the second-to-last column ($103 \leqslant Z \leqslant 118$), only approximately 14.6% of the nuclides involved in calculating the α decay energies for these 95 nuclides were included in the training set. Nevertheless, the RMSD values of the machine-learning-refined mass model decreased by approximately 7.5%, 8.7%, 8.9%, and 15.1% compared to the uncorrected model. This reduction further demonstrates the reliability of using machine learning to optimize nuclear mass models. To further validate this approach, the last column of the table presents the RMSD between the α decay energies and experimental values for 81 nuclides excluded from the training set. The data show that the machine-learning-refined mass model yields improvements to varying degrees compared to the pre-correction model. Furthermore, the last row of the table indicates that the results obtained using the BMA method are lower than those from other mass models. This finding suggests that the BMA method offers advantages in predicting nuclear masses.

**Table 3** The RMSD of the α-decay energy given by different mass tables and the experimental value (Unit: MeV)

| Model | $89 \leqslant Z \leqslant 118$ $^{203}$Ac–$^{294}$Og 371 nuclides | $89 \leqslant Z \leqslant 102$ $^{203}$Ac–$^{264}$No 276 nuclides | $103 \leqslant Z \leqslant 118$ $^{251}$Lr–$^{294}$Og 95 nuclides | $103 \leqslant Z \leqslant 118$ $^{251}$Lr–$^{294}$Og Nuclides outside the training set |
|---|---|---|---|---|
| LDM | 0.856 | 0.946 | 0.510 | 0.540 |
| DZ | 0.510 | 0.331 | 0.836 | 0.838 |
| WS4 | 0.253 | 0.238 | 0.291 | 0.307 |
| FRDM | 0.366 | 0.253 | 0.581 | 0.615 |
| LDM_lgb | 0.230 | 0.091 | 0.427 | 0.459 |
| DZ_lgb | 0.419 | 0.189 | 0.763 | 0.800 |
| WS4_lgb | 0.153 | 0.086 | 0.265 | 0.286 |
| FRDM_lgb | 0.263 | 0.098 | 0.493 | 0.530 |
| BMA | 0.135 | 0.079 | 0.230 | 0.245 |

Figure 4 illustrates the variation of $\alpha$-decay energy with neutron number $N$ for isotopes from Am to Og. The eight subplots respectively present the α-decay energies calculated using four mass models, comparing results before and after machine learning optimization against experimental measurements. The top four subplots correspond to even-Z nuclei, while the bottom four correspond to odd-Z nuclei. The same elements are selected for each model, with labels shown only for the WS4 model. The results from the machine learning-optimized mass tables align more closely with experimental values. For instance, the experimental α-decay energy of the $^{294}$Og nucleus is 11.874 MeV. The WS4 model optimized via machine learning yields 12.017 MeV, and the BMA model yields 12.092 MeV. Both values are closer to the experimental result than the 12.198 MeV predicted by the original WS4 mass table.

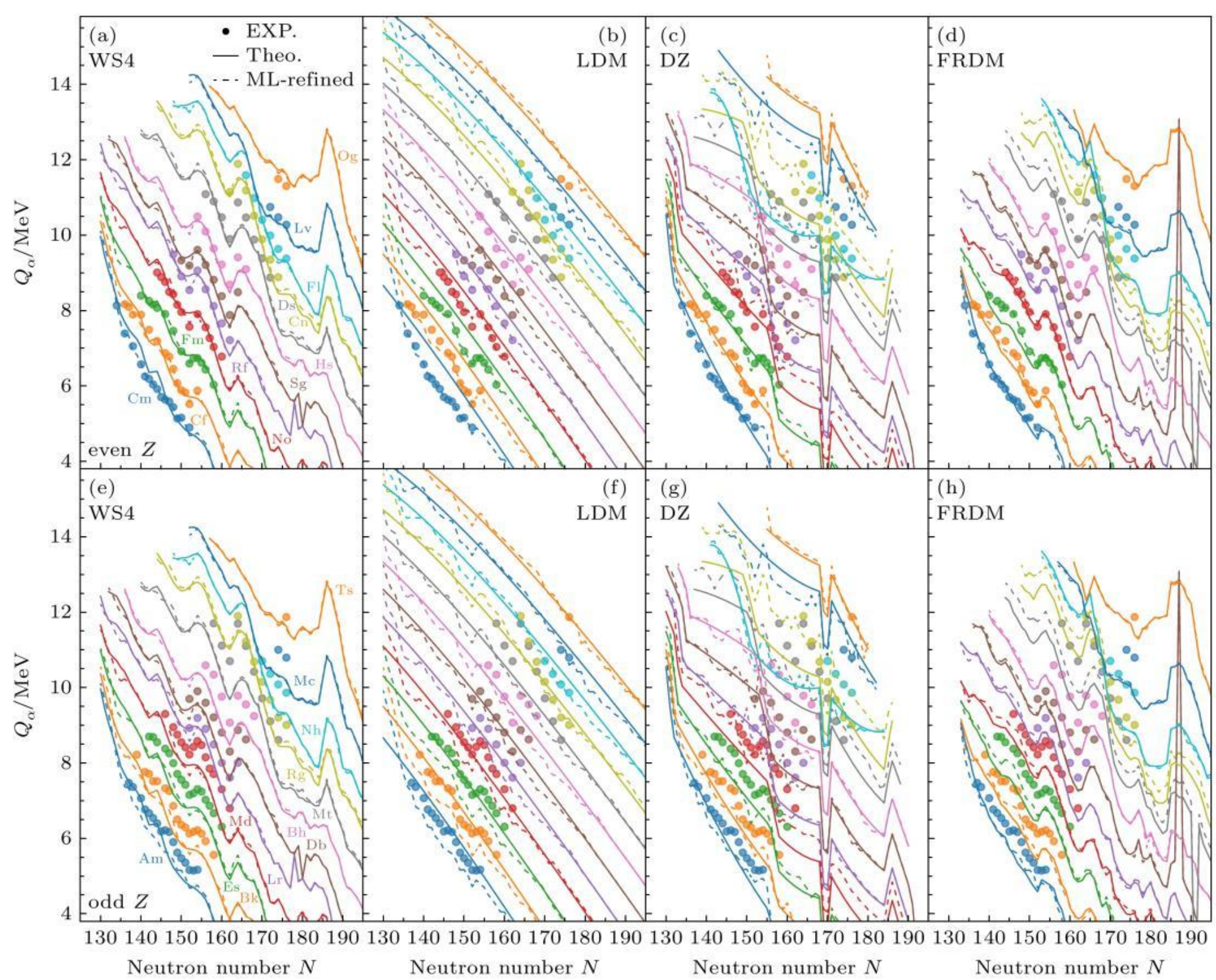

**Fig. 4 The variation of α-decay energy from Am to Og isotopes with neutron number N in the four mass model. The colored solid dots represent experimental values, the solid lines indicate theoretical values of the mass model, and the dashed lines show the results after machine learning optimization: (a)–(d) α-decay energy of even *Z* nuclei; (e)–(h) α-decay energy of odd *Z* nuclei.**

Currently, numerous laboratories worldwide are focusing on synthesizing elements with $Z = 119$ and $Z = 120$. Consequently, accurately predicting the $Q_\alpha$ values for the corresponding nuclides is essential. Figure 5 illustrates the variation in $Q_\alpha$ *values for* $Z = 119$ and 120 nuclides as a function of neutron number, according to various mass models. In Reference [61], the $Q_\alpha$ values of nuclei with existing experimental data were incorporated into the parameter fitting of an extended Skyrme energy density functional. This approach enables high-precision calculation of nuclear $Q_\alpha$ values, and the relevant results are also presented in Figure 5. Significant discrepancies exist among the results from different models. For instance, at a neutron number of approximately 170, the $Q_\alpha$ values predicted by various mass models differ by up to 1 MeV. This discrepancy leads to a difference of 1—2 orders of magnitude in the predicted α-decay half-lives[79,80]. Due to the scarcity of experimental data in the superheavy region, the results from the machine-learning-optimized mass table remain very close to those obtained before optimization.

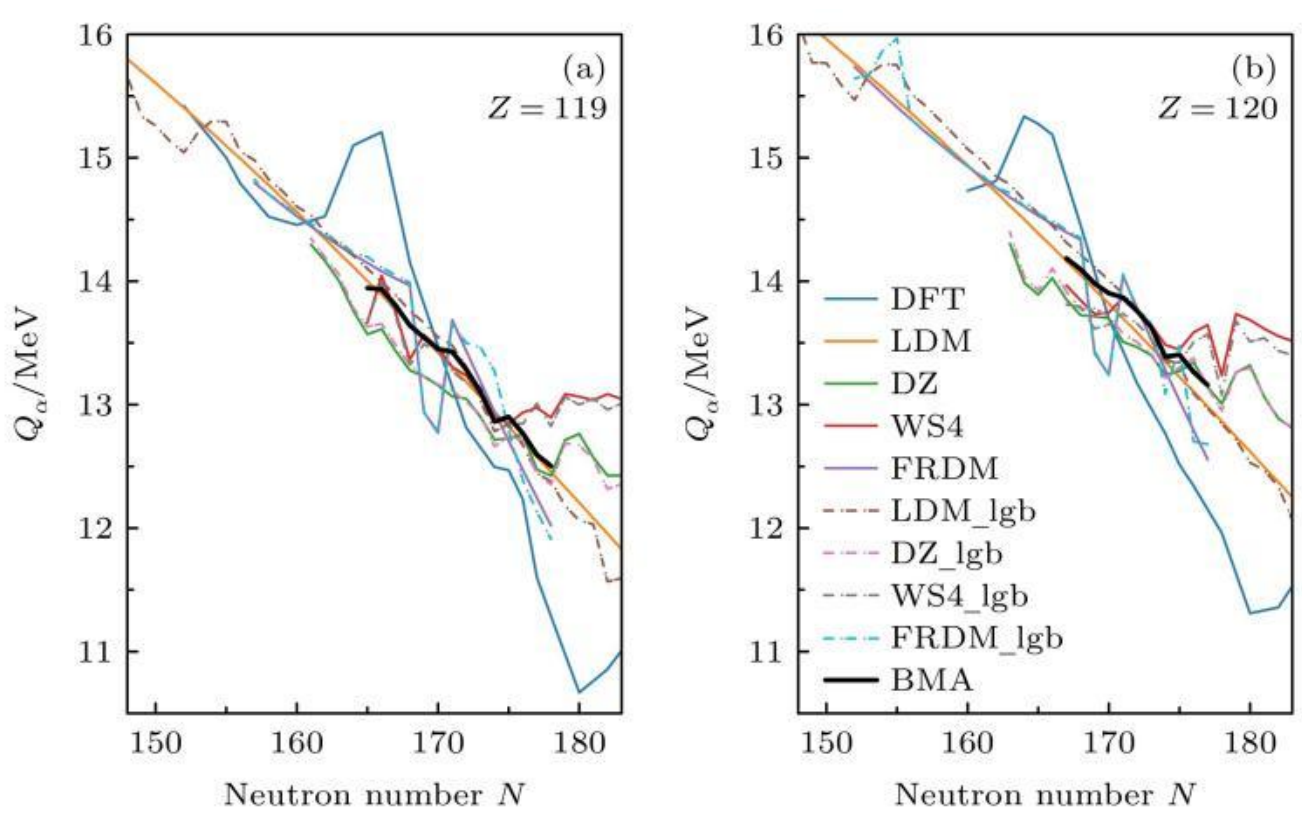

**Fig. 5 The variation of $Q_\alpha$ along the $Z$ = 119 and $Z$ = 120 isotopic chains with the change in neutron number. The solid curves in the figure represent the predicted values from the four mass models, while the dashed curves represent the results given by the LightGBM-refined mass model. BMA stands for the results of BMA, and DFT refers to the results presented in the Ref. [61].**

The experimental team at the Institute of Modern Physics, Chinese Academy of Sciences, is planning to synthesize element 119 using $^{54}$Cr as the projectile and $^{243}$Am as the target[81]. Based on current theoretical predictions[82], the synthesis probability is relatively high for the nucleus with neutron number 178 and mass number 297. The theoretical $Q_\alpha$ values for this nucleus vary across different mass models: 12.454 MeV (LDM), 12.020 MeV (DZ), 12.424 MeV (WS4), and 12.896 MeV (FRDM). After machine learning correction, the predicted $Q_\alpha$ values are 12.377 MeV (LDM-lgb), 11.911 MeV (DZ-lgb), 12.353 MeV (WS4-lgb), and 12.824 MeV (FRDM-lgb), respectively. The result based on the BMA model is 12.503 MeV. These results differ somewhat from the 11.286 MeV predicted in Reference [61]. In the coming years, the construction and operation of multiple large-scale heavy-ion nuclear physics facilities will accumulate increasing data in the superheavy nucleus region. These data will optimize nuclear mass models, thereby laying the foundation for accurately predicting the $Q_\alpha$ of element 119.

# 4 Conclusion

This study evaluated machine-learning-optimized mass tables using newly measured nuclear mass data since 2022, residual proton-neutron interactions, and $\alpha$-decay energies of heavy nuclei. The mean deviations of the machine-learning-optimized mass tables were smaller than those of the original tables, demonstrating the superiority of machine learning models in predicting nuclear masses. For 23 newly measured nuclei, the RMSD of the LDM, DZ, WS4, and FRDM mass tables optimized by LightGBM were 0.570, 0.558, 0.581, and 0.512 MeV, respectively. These values are lower than the pre-optimization values of 3.275, 1.058, 0.752, and 0.785 MeV. The RMSD based on the Bayesian Model Averaging method was 0.480 MeV, which is also lower than the results from the four machine-learning-optimized mass tables. The $\delta V_{pn}$ calculated using the machine-learning-optimized mass tables reproduces the anomalous bifurcation phenomenon of $\delta V_{pn}$ in $Z = N$ nuclei. It also predicts a peak in $\delta V_{pn}$ at $^{82}Nb$. Future experimental measurements of these nuclei will provide further opportunities to validate various nuclear mass tables. Regarding the prediction of $Q_\alpha$ for heavy nuclei, particularly for nuclides with $89 \leq Z \leq 102$, the RMSD of α-decay energies from the machine-learning-optimized mass tables can be less than 100 keV. This precision is valuable for experimental studies in this region. For nuclides with $103 \leqslant Z \leqslant 118$, the RMSD values of the machine-learning-corrected mass models decreased by approximately 7.5%, 8.7%, 8.9%, and 15.1% compared to the uncorrected models. This demonstrates the reliability of machine learning, enabling its broad application in analyzing larger nuclear datasets.

## Data Availability Statement

The dataset supporting the findings of this study is available in the Science Data Bank at https://doi.org/10.57760/sciencedb.j00213.00246.